\documentstyle[epsfig,preprint,epsf,aps]{revtex}
\tightenlines

\newcommand{\lsim}{\mathrel{\lower4pt\hbox{$\sim$}}
\hskip-12.5pt\raise1.6pt\hbox{$<$}\;}
\newcommand{\gsim}{\mathrel{\lower4pt\hbox{$\sim$}}
\hskip-12.5pt\raise1.6pt\hbox{$>$}\;}
 
\begin{document}
\baselineskip14pt

\tightenlines
 
\vspace{-.5in}                                                                           
\hskip4.5in
\vbox{\hbox{BNL--HET--00/18}
\hbox{hep-ph/0006280}}

\vspace{.5in}
\begin{center}
{\Large\bf Direct CP violation in radiative $b$ decays in and 
\\ \medskip beyond the Standard Model}                                                                                
\vspace{.5in}
 
Ken Kiers$^a$\footnote{Email: knkiers@tayloru.edu,
$^\dag$soni@bnl.gov, $^\ddag$wu@dirac.uoregon.edu}, Amarjit
Soni$^{b\dag}$ and Guo-Hong Wu$^{c\ddag}$
\vspace{.3in}
 
{\it $^a$Physics Department, Taylor University\\236 West Reade Ave., Upland,
IN 46989}
\medskip
 
{\it $^b$High Energy Theory, Department of Physics\\
Brookhaven National Laboratory,Upton, NY 11973-5000 }
\medskip
 
{\it $^c$Institute of Theoretical Science, University of Oregon, Eugene, OR 97403-5203}     
\end{center}
\bigskip
 
\baselineskip24pt                                                                        
\begin{abstract} 
We consider the partial rate asymmetry in the
inclusive decay modes $b \to s \gamma$ and $b \to d \gamma$,
concentrating on non-standard models with new
charged Higgs interactions.
We find that the charged Higgs contribution to
the asymmetry for $b \to s \gamma$ is small 
in such models due to a universal cancellation mechanism.
The asymmetry is therefore difficult to distinguish experimentally
from the Standard Model (SM) value, which is also small.
The cancellation mechanism 
is found to be rendered inoperative in supersymmetry
due to the presence of chargino loops.
Unlike $b \to s \gamma$, 
the rate asymmetry for $b \to d \gamma$ in Higgs models 
can be quite different from
its SM value, generally ranging from $-20\%$ to $+20\%$.
Specific model calculations are performed for the 
Three-Higgs Doublet Model and the ``Top'' Two-Higgs
Doublet Model to serve as illustrations.
We also offer some suggestions that may be helpful to experimentalists
in the detection of the inclusive mode $b\to d \gamma$.

\end{abstract}


\newpage
\tightenlines

\section{Introduction}

The $e^+ e^-$-based $B$-factories
are all performing quite well.  Each of them should soon be 
producing of order $10^7$ $B$-$\overline{B}$ pairs per year in a very clean
environment. In another few years it is likely that
one or more of these machines will
yield of order $10^8$ $B$-$\overline{B}$ pairs per year. 
$B$-experiments at the hadron machines --
Tevatron (BTEV), Hera-B and LHC-B --
could increase this number by another 1-3 orders of magnitude.
Therefore, studies of rare $B$-decays will continue
to intensify for the next several years.

Radiative $B$ decays have been the subject of much theoretical
and experimental interest over the past several years.  While
$b \to s \gamma$ is loop-suppressed within the SM, 
it nevertheless
has a relatively large branching ratio,
allowing for a very important comparison between theory and experiment.
A recent measurement of the branching ratio by CLEO~\cite{cleo1}
yielded the value ${\cal B}(b \to s \gamma) = (3.15 \pm 0.35 \pm 0.32 \pm 0.26)
\times 10^{-4}$, in good agreement with the
ALEPH measurement~\cite{aleph} and the 
SM prediction~\cite{chetyrkin,czarmarc,kagneub2}. 
The good agreement between the experimental and theoretical branching ratios
places a strong constraint on many non-standard models.
The recent CLEO measurements~\cite{cleo1} have
also placed a rather modest bound on the partial rate asymmetry (PRA) in 
$b \to s \gamma$, yielding $-9\%<A_{CP}^{b \to s \gamma}<42\%$ at the
90$\%$ confidence level.
This bound does not yet encroach on the very small PRA (of order 
$0.6\%$) predicted by
the SM~\cite{smpra}.  Improved measurements of $A_{CP}^{b \to s \gamma}$
are expected over the next several years at the $B$-factories.
These measurements should provide a powerful probe of new physics, 
particularly of models that contain non-standard CP-odd phases.

A mode closely related to $b\to s\gamma$ that is also of great
interest is $b\to d\gamma$.  The branching
fraction for this mode is expected to 
be of order $10^{-5}$, so a good
measurement of the branching fraction should
be possible with the $10^8$ $B$-$\overline{B}$ pairs per year expected
at the $B$-factories, provided the difficulties
in the experimental detection of $b\to d \gamma$ can be overcome (see below).  
The PRA for $b \to d \gamma$ is of order
$-16\%$ within the SM and may also be observable within the next several 
years.  It is important to note that 
the SM PRA for $b \to d \gamma$ is 
{\em statistically} more accessible 
than that for $b\to s \gamma$.
That is, the increase in the PRA for $b\to d\gamma$ 
relative to that for $b\to s \gamma$ more than compensates
the suppression of its branching ratio.  This statement
is relatively robust and holds true even after making reasonable
concessions for detection efficiencies in the two cases, as
we discuss in Sec.~\ref{sec:prainbdgamma}.  The statistical accessibility
of $A_{CP}^{b\to d \gamma}$ is quite intriguing and
leads us to a serious consideration of this observable in the present
paper.  A discussion of some practical experimental techniques 
for distinguishing
$b \to d \gamma$ from $b \to s \gamma$ 
may be found in Sec.~\ref{sec:exptech}.

Our goal in the present work is to assess the sensitivity
of the rate asymmetries noted above
to effects coming from various non-standard models.  
Our main emphasis is on effects
due to the exchange of charged Higgs bosons, but we
offer insights regarding other models as well.
A summary of our main results as well as those of some
previous authors may be found in Table~\ref{tableone}.
Previous studies of the $b \to s \gamma$ PRA due to charged Higgs exchange
have generically found quite small values~\cite{prahiggs,asatrian2,kagneub,BorzumatiG}.
In this paper we will show that this trend is due to
a partial cancellation between terms in a general expression
for the rate asymmetry.  The cancellation is a nearly
universal feature of models containing only new charged Higgs
interactions, and typically constrains $b \to s \gamma$ rate
asymmetries in such models to
be at or below the few-percent level.  Experimentally, such values would be
difficult to distinguish from the (similarly small) value expected
within the SM.  A contrasting situation is found for $b \to d \gamma$~\cite{asatrian2}.
In this case the PRA can differ appreciably from
the SM result, in general ranging between $-20\%$ and $+20\%$.

The outline of the remainder of the paper is as follows.  In 
Sec.~\ref{sec:formalism} we review the general formalism
used to calculate the PRA for inclusive radiative $b$ decays.
In Sec.~\ref{sec:higgs} we focus specifically on Higgs
models and show that $A_{CP}^{b \to s \gamma}$ is destined to be
quite small in such models, while $A_{CP}^{b \to d \gamma}$
can be both large and different from its SM value.  These general features
are illustrated by calculations in a few specific Higgs models, namely in
the Three-Higgs Doublet Model (3HDM) with Natural Flavour Conservation
and the Top-Two-Higgs Doublet Model (T2HDM).  Section~\ref{sec:susy} contains
a general discussion of supersymmetric effects on the asymmetries
and shows how supersymmetric models manage to evade the accidental
cancellation that Higgs models face.  Section~\ref{sec:exptech}
contains a discussion of some practical experimental
techniques that could be useful
in distinguishing between $b \to d \gamma$ and $b \to s \gamma$.
In Sec.~\ref{sec:concl} we offer a 
few concluding remarks, as well as a summary of our results.

\section{General formalism}
\label{sec:formalism}

Let us begin by reviewing the general formalism used to
calculate the PRA in 
inclusive radiative $B$ decays.  Further details of the calculation
may be found in Ref.~\cite{kagneub}.
The PRA in the inclusive decay $b \to s \gamma$ is defined by the ratio
\begin{eqnarray}
	A_{CP}^{b \to s \gamma} = \frac{
		\Gamma(\overline{B}\to X_s\gamma) - 
		\Gamma(B\to X_{\overline{s}}\gamma)}{
		\Gamma(\overline{B}\to X_s\gamma) +
                \Gamma(B\to X_{\overline{s}}\gamma)} \; .
	\label{acpbsg}
\end{eqnarray}
The next-to-leading order expression for the asymmetry 
may be found in Ref.~\cite{kagneub} and is expressed
in terms of the Wilson coefficients $C_i$ as
\begin{eqnarray}
	A_{CP}^{b \to s \gamma} & = & \frac{\alpha_s(m_b)}{
		|C_7|^2}\left\{
		\frac{40}{81} {\rm Im}\left[C_2C_7^*\right]
		-\frac{8z}{9}\left[v(z)+b(z,\delta_\gamma)\right]
		{\rm Im}\left[(1+\epsilon_s)C_2C_7^*\right] 
		\right. \nonumber \\
		& & \nonumber \\
		& &~~~~~~~~~~~~ \left.
		-\frac{4}{9}{\rm Im}\left[C_8C_7^*\right]
		+\frac{8z}{27} b(z,\delta_\gamma){\rm Im}\left[
		(1+\epsilon_s)C_2C_8^*\right] \right\} ,	
	\label{acpgen}
\end{eqnarray}
where $z=m_c^2/m_b^2\simeq (0.29)^2$ and
\begin{eqnarray}
	\epsilon_s = \frac{V_{us}^*V_{ub}}{V_{ts}^*V_{tb}} \simeq
		-\lambda^2(\rho-i\eta) .
	\label{eq:epss}
\end{eqnarray}
The second term in Eq.~(\ref{eq:epss}) gives an approximation
to $\epsilon_s$, written in the usual Wolfenstein parameterization,
with $\lambda\simeq 0.22$.
The functions $v(z)$ and $b(z,\delta_\gamma)$ are defined in Ref.~\cite{kagneub};
$\delta_\gamma$ is related to the experimental cut on the photon energy:
$E_\gamma > (1-\delta_\gamma) E_\gamma^{\rm max}$.  An analogous CP asymmetry can
be constructed for the decay $b \to d \gamma$ and gives an expression
similar to Eq.~(\ref{acpgen}), but with the replacement 
$\epsilon_s \to \epsilon_d$, where~\cite{kagneub}
\begin{eqnarray}
        \epsilon_d = \frac{V_{ud}^*V_{ub}}{V_{td}^*V_{tb}} \simeq
	\frac{\rho -i\eta}{1-\rho+i\eta} .
	\label{eq:epsd}
\end{eqnarray}
Kagan and Neubert have evaluated the coefficients of the
various terms in Eq.~(\ref{acpgen}) for different values of
the photon energy cut-off, $\delta_\gamma$.  They have also performed a
more elaborate calculation to take into account the effects of Fermi
motion~\cite{kagneub}.  For our purposes it is sufficient to neglect the Fermi-motion
effects and to choose a particular value of $\delta_\gamma$.  
Choosing $\delta_\gamma=0.30$,
and taking $\alpha_s(m_b)\simeq 0.214$, we find
\begin{eqnarray}
       A_{CP}^{b \to s(d) \gamma} & \simeq & \frac{10^{-2}}{
                |C_7|^2}\biggl\{ 1.17 \times {\rm Im}\left[C_2C_7^*\right] 
                - 9.51 \times {\rm Im}\left[C_8C_7^*\right] 
		+0.12\times {\rm Im}\left[C_2C_8^*\right] \biggr.
		\nonumber \\
		& & ~~~~~~~~~ \biggl.
		-9.40 \times {\rm Im}\left[\epsilon_{s(d)}C_2\left(
		C_7^*-0.013 \; C_8^*\right)\right]\biggr\} .
	\label{acpgen2}
\end{eqnarray}
The rather large coefficient in the second term
has led to the observation that models with enhanced
chromomagnetic dipole operators could give rise to 
significant changes in the rate asymmetries, compared to those of the SM~\cite{kagneub}.

The Wilson coefficients in Eq.~(\ref{acpgen2}) are
determined using the usual Renormalization Group Evolution (RGE)
procedure~\cite{bbl} and are to be evaluated
at the $b$ quark mass scale.  The results at leading order are~\cite{c7note}
\begin{eqnarray}
	C_2(m_b) & = & \sum_{i=1}^8 k_{2i}\tilde{\eta}^{a_i} 
 	 \label{eq:c2} \\
	C_7(m_b) & = & \tilde{\eta}^{16/23}C_7(m_W) +
		\frac{8}{3}\left(\tilde{\eta}^{14/23}-
			\tilde{\eta}^{16/23}\right)
		C_8(m_W) + C_2(m_W)\sum_{i=1}^8 h_i\tilde{\eta}^{a_i} 
	 \label{eq:c7} \\
	C_8(m_b) & = & \tilde{\eta}^{14/23}C_8(m_W) +        
           	C_2(m_W)\sum_{i=1}^8\overline{h}_i\tilde{\eta}^{a_i},
	 \label{eq:c8}
\end{eqnarray}
where $\tilde{\eta}\equiv \alpha_s(m_W)/\alpha_s(m_b)$.  The coefficients
$k_{2i}$, $h_i$ and $\overline{h}_i$ are tabulated in Ref.~\cite{bbl}.
The SM expressions for the Wilson coefficients are given by
\begin{eqnarray}
        C_2^{\rm SM}(m_W) & = & 1
         \label{eq:c2b} \\
        C_7^{\rm SM}(m_W) & = & -\frac{1}{2} A(x_t)
         \label{eq:c7b} \\
        C_8^{\rm SM}(m_W) & = & -\frac{1}{2} D(x_t) ,
         \label{eq:c8b}
\end{eqnarray}
where $x_t\equiv m_t^2/m_W^2$ and where
$A$ and $D$ are the well-known ``Inami-Lim'' functions~\cite{inamilim}:
\begin{eqnarray}
	A(x) & = & x\left[\frac{8x^2+5x-7}{12(x-1)^3} - \frac{(3x^2-2x)\ln x}
		{2(x-1)^4}\right]
	 \label{eq:acap} \\
	D(x) & = & x\left[\frac{x^2-5x-2}{4(x-1)^3}+\frac{3x\ln x}
		{2(x-1)^4}\right]
	 \label{eq:dcap} .
\end{eqnarray}

Deviations from the SM may be incorporated into the Wilson coefficients
at the $W$ mass scale and can then be 
run down to the $b$ mass scale using the RGE
equations given in Eqs.~(\ref{eq:c2})-(\ref{eq:c8}).  This procedure yields the
following simple expressions for $C_7(m_b)$ and 
$C_8(m_b)$~\cite{kagneub},
\begin{eqnarray}
	C_7(m_b) & \simeq & -0.31 + 0.67\; C^{\rm new}_7(m_W)
		+ 0.09\; C^{\rm new}_8(m_W) 
	\label{eq:c7eff} \\
	C_8(m_b) & \simeq & -0.15 + 0.70\; C^{\rm new}_8(m_W),
	\label{eq:c8eff}
\end{eqnarray}
where we have used $\tilde{\eta}\simeq 0.56$ 
(corresponding to $\alpha_s(m_Z)=0.118$)
and where the Wilson coefficients at the $W$ mass scale are given by
\begin{eqnarray}
        C_7(m_W) & = & C_7^{\rm SM}(m_W) + C_7^{\rm new}(m_W) \\
	C_8(m_W) & = & C_8^{\rm SM}(m_W) + C_8^{\rm new}(m_W) .
\end{eqnarray}
Inserting Eqs.~(\ref{eq:c7eff}) and (\ref{eq:c8eff}) into Eq.~(\ref{acpgen2})
yields the following general expression for the CP asymmetry
\begin{eqnarray}
       A_{CP}^{b \to s(d) \gamma} & \simeq & \frac{10^{-2}}{
                |C_7(m_b)|^2}\biggl\{ -1.82 \times {\rm Im}\left[C_7^{\rm new}\right] 
                +1.72 \times {\rm Im}\left[C_8^{\rm new}\right] 
		-4.46\times {\rm Im}\left[C_8^{\rm new}
			C_7^{{\rm new} *}\right] 
			\biggr.
		\nonumber \\
		& & ~~~~~~~~~ \biggl.
		+3.21 \times {\rm Im}\left[\epsilon_{s(d)}\left(
		1-2.18 \; C_7^{{\rm new} *}
		 -0.26\; C_8^{{\rm new} *}\right)\right]\biggr\} .
	\label{acpgen3}
\end{eqnarray}	
Note that if $C_7^{\rm new}$ and $C_8^{\rm new}$ are approximately
equal in a particular scenario, 
then the third term in the above expression is close to zero, 
and the first two terms nearly cancel one another.
The main contribution to the PRA in that case is the last
term.  Such a scenario yields a rather small
asymmetry for $b \to s \gamma$ (since the asymmetry is essentially 
proportional to $\epsilon_s$), but possibly a sizable asymmetry
for $b \to d \gamma$.
As we shall see below, $C_7^{\rm new}\approx C_8^{\rm new}$ in multi-Higgs
doublet models, so that $b\to s \gamma$ generically has a small
rate asymmetry in such models.

Setting both $C_7^{\rm new}(m_W)$ and $C_8^{\rm new}(m_W)$ to zero
in the above expression 
yields the SM predictions for the CP-violating
rate asymmetries in $b \to s \gamma$ and $b \to d \gamma$,
\begin{eqnarray}
 	A_{CP}^{b \to s(d) \gamma} & \simeq & 0.334 \times {\rm Im}
		\left[\epsilon_{s(d)}\right],
\end{eqnarray}
or, in terms of the angles of the familiar unitarity triangle~\cite{note:beta_CKM},
\begin{eqnarray}
	A_{CP}^{b \to s \gamma} \simeq 0.334 \times \lambda^2 \; \frac{\sin\beta_{CKM}\sin\gamma}
	{\sin\alpha}\; , ~~~~~~~~ 
	A_{CP}^{b \to d \gamma} \simeq - 0.334 \times \frac{\sin\alpha\sin\beta_{CKM}}
	{\sin\gamma}\; .
\end{eqnarray}
Substituting the approximate expressions for $\epsilon_s$ and $\epsilon_d$
from Eqs.~(\ref{eq:epss}) and (\ref{eq:epsd}) and using the ``best-fit''
values $(\rho,\eta)=(0.20,0.37)$~\cite{alilondon}, we find
\begin{eqnarray}
	\left. A_{CP}^{b \to s \gamma}\right|_{SM} = 0.6\% \;, ~~~~~~~~~~
	\left. A_{CP}^{b \to d \gamma}\right|_{SM} = -16\% \; ,\nonumber
\end{eqnarray}
in agreement with the estimates of previous authors~\cite{kagneub}.
The rate asymmetry for $b \to s \gamma$ is potentially a sensitive probe
of non-standard physics, since its value within the SM is so small.  Finding an
experimental value larger than a few percent would be 
a strong indication of new physics.  For $b\to d \gamma$, the rate
asymmetry is already large within the SM; in this case,
contributions due to non-standard
physics might be more difficult to disentangle from the SM contribution
unless they lead to an appreciable change in the PRA.

\section{Multi-Higgs doublet models}
\label{sec:higgs}

One of the simplest ways to extend the SM is to expand its Higgs sector by
adding extra Higgs doublets.  Despite their apparent simplicity, Multi-Higgs doublet 
models can have rich phenomenological consequences.  Of particular interest within
the context of the present work are those models that give rise to new sources of CP violation.
If one imposes Natural Flavour Conservation (NFC)~\cite{NFC}, 
CP violation in the charged Higgs sector
does not appear until two extra Higgs doublets have been added.  The Three-Higgs Doublet
Model (3HDM) with NFC, incorporating spontaneous and/or explicit
CP violation, has been studied extensively over the last several
decades~\cite{weinberg3hdm,chang,atwood,grossman,grnir}
and in many instances it leads to large CP-violating effects.
Another option is to relax the requirement
of NFC.  In this case, CP violation appears with only one extra Higgs doublet.  The so-called
``Type-III'' Two-Higgs doublet models can lead to interesting phenomenological
consequences~\cite{type3}.  One variant on this class of models -- the ``Top-Two-Higgs 
Doublet Model'' (T2HDM) -- gives special status to
the top quark~\cite{daskao,ksw}.  The T2HDM has couplings
unlike most other Higgs models and can produce significant effects on CP-odd observables
in decays such as $B\to\psi K_S$ and $B^\pm \to\psi K^\pm$ and in $D$-$\overline{D}$
mixing~\cite{daskao,ksw,zhou}.

Bearing in mind the above comments, it would seem natural to assume that Multi-Higgs doublet
models could give rise to significant effects in the rate asymmetries for radiative $b$
decays.  This intuitive guess is correct for $b \to d \gamma$, but is {\em not} correct
for $b \to s \gamma$.  Small values for
$A_{CP}^{b\to s \gamma}$ were reported in 
previous studies~\cite{prahiggs,asatrian2,kagneub,BorzumatiG}, but the reason
for the smallness of the asymmetry was not elucidated.  In the following we show
that the rate asymmetry for $b \to s \gamma$ is typically less than a few percent
in Multi-Higgs doublet models.  The analogous asymmetry for $b \to d \gamma$ can 
nevertheless be strongly affected by the presence of extra Higgs doublets.  Our general
arguments will be supported by model calculations in the 3HDM and T2HDM.

\begin{figure}[hbt]
\epsfxsize5in
\epsfbox{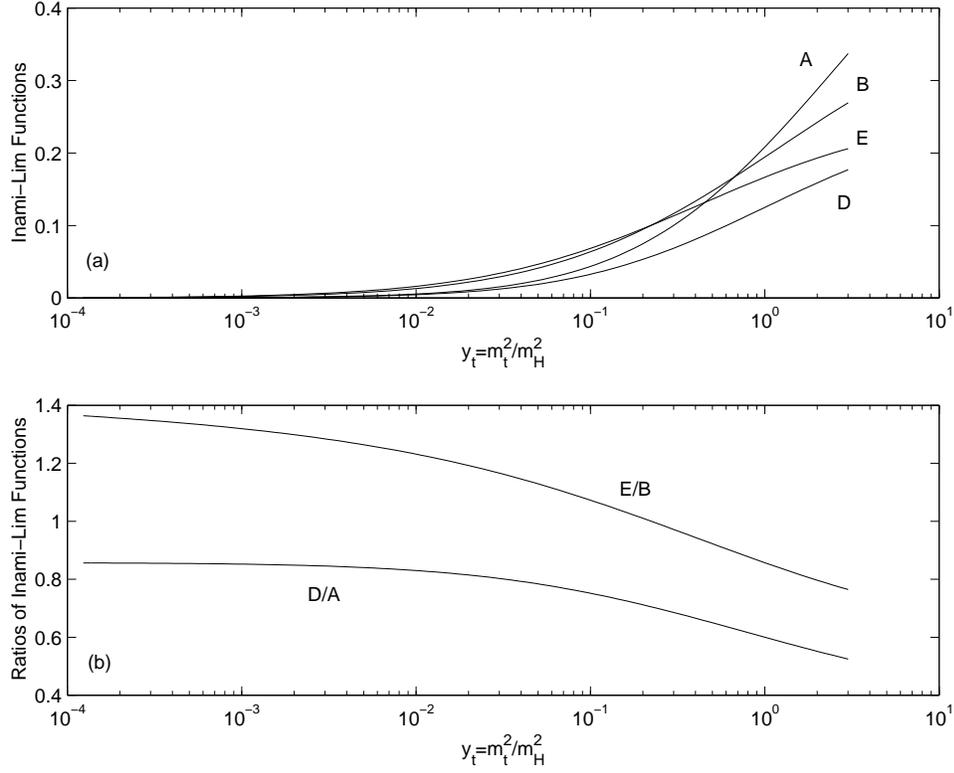}
\medskip
\medskip
\caption{(a) The Inami-Lim functions as defined in the text.  (b) Ratios of
the Inami-Lim functions.}
\label{fig:inami}
\end{figure}

The rate asymmetry for $b \to s \gamma$ is small in Higgs models
because $C_8^{\rm new}(m_W)/C_7^{\rm new}(m_W)$$\sim$$1$ in these
models.  As noted in the discussion following Eq.~(\ref{acpgen3}), this approximate
equality leads to a strong suppression of the rate asymmetry.
If we only consider contributions to the rate asymmetry
arising from diagrams containing a charged Higgs and a top quark in the
loop~\cite{fcnhnote},
the ``new physics'' corrections to $C_7$ and $C_8$ have the following
general form
\begin{eqnarray}
	C_7^{\rm new}(m_W) & = & f_A A(y_t) + f_B B(y_t) \\
	C_8^{\rm new}(m_W) & = & f_A D(y_t) + f_B E(y_t) ,
\end{eqnarray}
where $y_t\equiv m_t^2/m_H^2$, and $A(y)$ and $D(y)$ are the
Inami-Lim functions introduced above.  The functions $B(y)$ and $E(y)$
are given by
\begin{eqnarray}
  B(y) & = &  y \left[
      \frac{5y-3}{12(y-1)^2} -\frac{(3y-2)\ln y}{6(y-1)^3}
             \right] \\
  E(y) & = &  y \left[
      \frac{y-3}{4(y-1)^2} + \frac{\ln y}{2 (y-1)^3}
           \right] \; .
\end{eqnarray}
Figure~\ref{fig:inami}(a) shows the various Inami-Lim functions,
plotted as functions of the dimensionless variable $y_t$, while
Fig.~\ref{fig:inami}(b) shows plots of the ratios $D/A$ and $E/B$.
As is evident from these plots, $D/A$ and $E/B$ are
both of order unity for Higgs masses above $100$~GeV.  As a result,
$C_8^{\rm new}/C_7^{\rm new}$ is of order unity and
the CP asymmetry for $b\to s \gamma$
is expected to be rather small.
Note that the deviation from unity is slightly larger for $D/A$ than for $B/E$
when $m_H\sim m_t$.  Thus, slightly larger asymmetries may be expected 
in scenarios where $f_A\gg f_B$.  (See the footnote and discussion following
Eq.~(\ref{zeta3hdm}).)

Let us quantify the above argument by parametrizing the deviation of 
$C_8^{\rm new}/C_7^{\rm new}$ from unity in terms of a complex quantity
$\zeta$.  Defining
\begin{eqnarray}
 C_8^{\rm new}(m_W) & \equiv & 
(1+\zeta)  C_7^{\rm new}(m_W) \; ,
\end{eqnarray}
we have
\begin{eqnarray}
       A_{CP}^{b \to s(d) \gamma} & \simeq & \frac{10^{-2}}{
                |C_7(m_b)|^2}\biggl\{ 
			-0.10 \; {\rm Im}\left[C_7^{\rm new}\right] 
                +1.72 \; {\rm Im}\left[\zeta C_7^{\rm new}\right] 
		-4.46\left| C_7^{\rm new}\right|^2
			\; {\rm Im}\left[\zeta\right] 
			\biggr.
		\nonumber \\
		& & ~~~~~~~~~~~~~~~ \biggl.
		+3.21 \; {\rm Im}\left[\epsilon_{s(d)}\left(
		1-2.44 \; C_7^{{\rm new} *}
		 -0.26\; \zeta^* C_7^{{\rm new} *}\right)\right]\biggr\} .
	\label{acpgenhiggs}
\end{eqnarray}	
Each of the first three terms in the above expression is small, either because it is
suppressed by $\zeta$ or,
in the case of the first term, because it has suffered an accidental
cancellation.  Since $\epsilon_s$ is itself quite small,
the asymmetry for $b\to s \gamma$ will never be much larger than
several percent.  The present analysis is quite general and predicts that the CP asymmetry
in $b \to s \gamma$ will be quite small in Higgs models.
This result is independent of whether or not the corrections to the Wilson
coefficients have complex phases.
The situation for $b \to d \gamma$ is quite different.  While it is still
true that the first three terms in Eq.~(\ref{acpgenhiggs})
make only a small contribution to the asymmetry, the last term now
contains $\epsilon_d$, which is relatively large and can have a significant
effect on the asymmetry.

\subsection{Three-Higgs doublet model with NFC}

We now consider the 3HDM with explicit CP violation, since the 3HDM with
spontaneous CP violation alone is disfavoured by data~\cite{chang,grnir}.
The charged-Higgs Yukawa couplings for this model may be written as~\cite{weinberg3hdm}
\begin{eqnarray}
L_{H^+} & = & (2\sqrt{2} G_F)^{\frac{1}{2}} H^+ (
 X \overline{U}_L V M_D D_R
+Y \overline{U}_R M_U V D_L
+ Z \overline{N}_L M_E E_R )  + {\rm H.c.}
\end{eqnarray}
where $X$, $Y$, and $Z$ are complex parameters~\cite{lightnote}.
Note that the substitution
$X, Z \to \tan \beta$ and $Y \to \cot \beta$ yields the 
Yukawa interactions of the Type-II 2HDM.
The Wilson coefficients at the $m_W$ scale are given by
\begin{eqnarray}
C_7(m_W) & = & - \frac{1}{2} A(x_t)
              - X Y^* B(y_t) -\frac{1}{6} |Y|^2 A(y_t) \nonumber \\
C_8(m_W) & = & - \frac{1}{2} D(x_t) 
           - X Y^* E(y_t) -\frac{1}{6} |Y|^2 D(y_t)
	\label{c7c83hdm} \; .
\end{eqnarray}
Thus 
\begin{eqnarray}
f_A & = &  -\frac{1}{6} |Y|^2 \nonumber \\
f_B & = & - X Y^* \nonumber \\
\zeta & = & \frac{X Y^* [ E(y_t) - B(y_t) ] 
                  + \frac{1}{6} |Y|^2 [ D(y_t) - A(y_t) ] }
        { X Y^* B(y_t)  + \frac{1}{6} |Y|^2 A(y_t) } \; .
	\label{zeta3hdm}
\end{eqnarray}
In the numerical work that follows, $|\zeta|$ is typically
found to be less than $0.20$, so Eq.~(\ref{acpgenhiggs}) yields
a small asymmetry for $b \to s \gamma$, as noted above~\cite{zetanote}.

The parameter space for this 3HDM depends on four independent
quantities, which may be taken to be $|X|$, $|Y|$, $\arg(XY^*)$
and $m_H$.  The phase of $Y$ may
be absorbed into the definition of $X$ without affecting
any observable in $b \to s(d) \gamma$.  We thus take $Y$ to
be real and non-negative and write $X$ as
\begin{eqnarray}
	X = X_R + i X_I,
\end{eqnarray}
where $X_R$ and $X_I$ are the real and imaginary parts of $X$.
Possible experimental constraints on $X$ and $Y$ come
from neutral kaon decays, the neutron electric dipole moment,
$B$-$\overline{B}$ mixing and the (CP-averaged)
branching ratio for $b\to s \gamma$~\cite{grossman,grnir,chang,hewett}.
Another constraint may be obtained by requiring the Higgs sector of
the theory to be perturbative~\cite{pert}.
Most of these considerations lead to fairly mild constraints
on the parameter space of the 3HDM.  Perturbativity requires
$|X|\lsim 130$~\cite{pert}, while $B$-$\overline{B}$ mixing
requires $|Y|\lsim 1$ for $m_H\sim 200$~GeV.  The neutron electric
dipole moment places a very weak constraint on Im$(XY^*)$~\cite{dn,babarphysics}.
A much stronger bound may be obtained from the branching ratio
for $b\to s \gamma$, which yields, for example, Im$(XY^*)\lsim 3.5$ 
for $m_H=200$~GeV (see Eq.~(\ref{eq:imxy}) below, or Fig.~\ref{fig:3hb}).
$\epsilon_K$ may also be used to constrain the parameter space
of this model.  If $\rho$ and $\eta$ take their SM central values,
the constraint on $Y$ is similar to that coming from $B$-$\overline{B}$
mixing.  Long distance effects due to charged Higgs exchange may also
be important for $\epsilon_K$~\cite{chang,donoghue}.  Given the 
theoretical uncertainties associated with long-distance
calculations, we ignore such effects in our analysis.

The strongest bounds on $X$ and $Y$ come from
the (CP-averaged) branching ratio for $b \to s \gamma$.  In order
to minimize uncertainties due to the mass of the bottom quark, it is common
to define the ratio
\begin{eqnarray}
	R = \frac{\Gamma(b \to s \gamma)}{\Gamma(b\to c e \nu)}.
\end{eqnarray}
Calculation of $R$ at leading order gives~\cite{ratebsg}
\begin{eqnarray}
	R\simeq 0.0231 \times |C_7(m_b)|^2 ,
		\label{eq:Rth}
\end{eqnarray}
where $C_7(m_b)$ is given in Eq.~(\ref{eq:c7eff}).  The region allowed
by the $b \to s \gamma$ branching ratio is bounded by two concentric
circles in the complex $X$-plane for given values of $Y$ and $m_H$.
This may be seen by writing 
\begin{eqnarray}
	C_7(m_b)=-c-Xd, 
\end{eqnarray}
where $c$ and $d$ depend only on $Y$ and $m_H$:
\begin{eqnarray}
	c & = & 0.31 + \frac{1}{6}\left[0.67 A(y_t)+0.09 D(y_t)\right]\left|Y\right|^2 
		\label{eq:c} \\
	d & = & \left[0.67 B(y_t)+0.09 E(y_t)\right]Y .
		\label{eq:d}
\end{eqnarray}
The constraint due to the measured rate for $b \to s \gamma$ then takes the
general form
\begin{eqnarray}
	R^{\rm min} \leq 0.0231 \times d^2 \left[\left(X_R+\frac{c}{d}\right)^2
		+X_I^2\right] \leq R^{\rm max};
\end{eqnarray}
that is, the allowed region is bounded by two concentric circles in
the complex $X$-plane.
$R^{\rm max}$ and $R^{\rm min}$ depend on the experimental
value for the ratio $R$ as well as on various experimental and theoretical 
uncertainties.  The above expression may be used to find an upper bound
on $|$Im$(XY^*)|$.  In terms of the geometrical picture presented above,
the upper bound occurs at the ``top'' and ``bottom'' of the outer circle, yielding
\begin{eqnarray}
	|\mbox{Im}(XY^*)| \leq \frac{1}{0.67 B(y_t)
		+ 0.09 E(y_t)} \sqrt{\frac{R^{\rm max}}{0.0231}} \; .
	\label{eq:imxy}
\end{eqnarray}

A search of the parameter space for this 3HDM yields results
consistent with the general arguments given for Higgs models at the
beginning of Sec.~\ref{sec:higgs}.
The rate asymmetry for $b \to s \gamma$ is found to be
smaller than three or four percent, while that for $b\to d \gamma$ can vary between
$\pm 20\%$.  Figures~2 and 3 show plots that are representative
of the results obtained.  In each figure, contours of constant
asymmetry for $b \to s(d) \gamma$ are shown
in the $(X_R,X_I)$ plane for fixed values of $Y$ and $m_H$.
Labels for the various contours are given as percentages.
The size of the ring-shaped allowed region is determined by
the maximum and minimum values allowed for the ratio $R$.
The current experimental value for this ratio 
may be determined by using the weighted average of
the CLEO~\cite{cleo1} and ALEPH~\cite{aleph} values
for the $b\to s \gamma$ branching ratio, 
${\cal B}(B\to X_s \gamma)|_{\rm ave} \simeq (3.14 \pm 0.48)\times 10^{-4}$,
as well as the branching ratio for $b\to c e \nu$, 
${\cal B}(B \to X_c e \nu) \simeq 0.104\pm 0.009$~\cite{pdg}.
This procedure yields the value $R = (3.02 \pm 0.53) \times 10^{-3}$.
In addition to the experimental uncertainty for this 
ratio, there is another theoretical uncertainty.
Since we are using LO expressions for the Wilson coefficients,
we estimate the magnitude of this uncertainty to be
$\delta R^{\rm theory} = 0.7 \times 10^{-3}$.  Combining the
experimental and theoretical uncertainties in quadrature
and doubling the result to get a ``$2\sigma$'' uncertainty,
we obtain $R^{\rm min} = 0.0013$ and $R^{\rm max} = 0.0048$.
These are the values used in the plots.

\begin{figure}[hbt]
\epsfxsize5in
\epsfbox{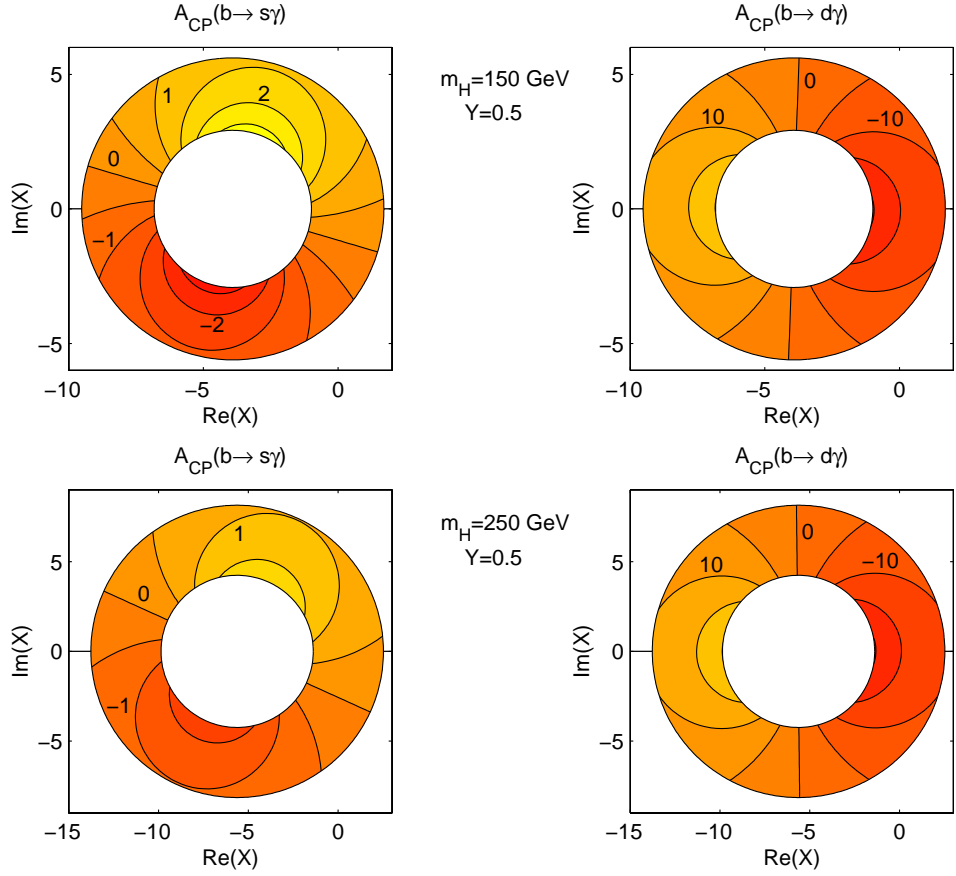}
\medskip
\medskip
\caption{
Contour plots of partial rate asymmetries in the decays
$b\to s \gamma$ and $b\to d \gamma$ in the 3HDM with
$\rho=0.20$ and $\eta=0.37$.  
The shaded ring-like regions are experimentally allowed. 
The contour lines show the CP asymmetries as percentages.}
\label{fig:3ha}
\end{figure}

Figure~\ref{fig:3ha} compares the asymmetries obtained for two different values
of the Higgs mass with $Y$ held constant, while Fig.~\ref{fig:3hb} keeps $m_H$
constant and shows the effect of changing $Y$.  
While the size of the allowed region exhibits a strong dependence on 
the values chosen for $Y$ and $m_H$ (note the changes in
the horizontal and vertical scales in the plots), the contours themselves
look very similar from plot to plot.  This is particularly true in the
case of $b \to d \gamma$, whose asymmetry contours 
appear to be exact replicas of one another once
an appropriate scaling and shifting has been performed in the $(X_R,X_I)$
plane.  In order to understand the scaling behaviour
of $A_{CP}^{b \to d \gamma}$, it is easiest to return to one of the
original rate asymmetry expressions, Eq.~(\ref{acpgen2}).  The 
first three terms in Eq.~(\ref{acpgen2})
are identical for $b \to s \gamma$ and $b \to d \gamma$
in this model.  According to the discussion following Eq.~(\ref{acpgenhiggs}), these
terms combine to give a contribution to the asymmetry of at most
a few percent.  A good approximation for $b \to d \gamma$
is thus obtained by neglecting these terms altogether (as well as
the small contribution proportional to $\epsilon_d C_2 C_8^*$) and writing
\begin{eqnarray}
       A_{CP}^{b \to d \gamma} & \simeq & \frac{-9.40\times 10^{-2}}{
                |C_7(m_b)|^2}\times {\rm Im}\left[\epsilon_{d}C_2 C_7^*\right].
		\label{eq:acpbdgappr}
\end{eqnarray}
Changes in $Y$ or $m_H$ lead to changes in the functions $c$ and $d$ defined
in Eqs.~(\ref{eq:c}) and (\ref{eq:d}).  It is always possible to perform a linear
transformation on $X_R$ and $X_I$ that compensates for this
change and leaves $C_7(m_b)$ invariant.
Since both the above approximate expression for $A_{CP}^{b \to d \gamma}$ 
and the expression for the rate, Eq.~(\ref{eq:Rth}), depend only
on $C_7(m_b)$, they too are invariant under this 
transformation~\cite{transnote}.
The variations in the contours for $b \to s \gamma$ may be attributed to
the fact that $C_8(m_b)$ -- which has a larger effect on $b \to s \gamma$ --
is not invariant under the transformation.

Figure~\ref{fig:3hc} shows a few representative correlation plots 
demonstrating the relationship between
the asymmetries for $b\to s \gamma$ and $b \to d\gamma$.  It also shows
the relationship between the asymmetry and branching ratio for $b \to d\gamma$.
The branching ratios in these plots
have been normalized to the theoretical expression 
for ${\cal B}(b \to s \gamma)$ within the SM.  The ratio
of branching ratios has the simple form 
\begin{eqnarray}
	\frac{{\cal B}_{\rm 3HDM}(b\to d \gamma)}{{\cal B}_{\rm SM}(b\to s \gamma)}
		= \left|\frac{V_{td}}{V_{ts}}\right|^2
			\left|\frac{C_7(m_b)|_{\rm 3HDM}}
			{C_7(m_b)|_{\rm SM}}\right|^2 .
\end{eqnarray}
The points for these plots are taken to be uniformly distributed on square 
grids inside the respective allowed regions in the complex $X$-plane.
This method of choosing points leads to the observed
``textures'' in these plots.

\subsection{Type-III Two-Higgs Doublet Models}

  In Two-Higgs Doublet Models with natural flavour conservation
\cite{NFC} (as in types I and II \cite{hhunter}), there exist no new CP-violating phases
in the charged-Higgs Yukawa couplings beyond that
of the Cabibbo-Kobayashi-Maskawa (CKM)~\cite{kobay} matrix.
Consequently, the charged Higgs contribution to the
$b \to s \gamma$ rate asymmetry is both CKM- and QCD-suppressed
(the latter by a factor $\alpha_s/\pi$)
and the asymmetry is small, as in the SM.  The $b\to d \gamma$ rate asymmetry is also similar
to its SM value of approximately $-16\%$.
In Type III 2HDM's \cite{type3},  where no ${\it ad hoc}$ discrete 
symmetry is imposed, one generally expects new CP-violating phases
in the Yukawa couplings (as well as flavour-changing neutral Higgs interactions).
As a result, the CKM and GIM suppressions present in the SM for $b\to s \gamma$
are in general
no longer operative, and a potentially large partial rate asymmetry
in $b \to s \gamma$ seems 
possible.   This expectation is invalid due to the
accidental cancellation mechanism discussed above.

As a concrete example of a Type-III 2HDM, we consider the
2HDM for the top quark (T2HDM) \cite{daskao,ksw}. 
This model is designed to accommodate the heaviness of the
top quark by coupling it to a scalar-doublet with a large VEV.  The
other five quarks are coupled to the other scalar-doublet, whose VEV
is much smaller. As a result, $\tan \beta$ is naturally large in this model.
A distinctive feature of the T2HDM is its $\tan \beta$-enhanced 
charm quark Yukawa coupling, with the consequence that 
the charged Higgs sector can significantly affect the neutral kaon 
system and the $b \to s \gamma$ rate.

Although there are two new CP phases in the T2HDM from
the unitary rotation of the right-handed up-type quarks, only
one of them contributes to radiative $b$ decays. 
This is the phase $\delta$ of the complex parameter $\xi=|\xi| e^{-i\delta}$,
measuring the $c_R$-$t_R$ mixing~\cite{ksw}. 
As was done before, we take $|\xi|=1$ throughout this paper.
In contrast with the 3HDM,  the virtual charm quark contribution
in the T2HDM
can occasionally be of the same magnitude as that of the top quark. 
The Wilson coefficient for $b\to q \gamma$ ($q=s,d$) is given by~\cite{ksw}
\begin{eqnarray}
        C^{\rm new}_7(m_W) & = & \sum_{i=c,t}\kappa^{iq}
                \left[ -\tan^2\beta +\frac{1}{m_i V_{iq}^\ast}
                \left(\Sigma^T V^\ast\right)_{iq}\left(\tan^2\beta + 1
                        \right)\right] \nonumber \\
                & & \times\left\{B(y_i) + \frac{1}{6}A(y_i)
                        \left[-1 + \frac{1}{m_i V_{ib}}
                \left(\Sigma^\dagger V\right)_{ib} \left(\cot^2\beta + 1
                        \right)\right]\right\} ,
        \label{eq:delc7}
\end{eqnarray}
where $\kappa^{iq} = -V_{ib}V^*_{iq}/(V_{tb}V^*_{tq})$,
$y_i = (m_i/m_H)^2$, and $V$ denotes the CKM 
matrix.
The definition of the matrix $\Sigma$ may be found in \cite{ksw}.
The expression for $C^{\rm new}_8(m_W)$ is obtained
from $C^{\rm new}_7(m_W)$ by the substitutions
$A(y_i) \rightarrow D(y_i)$ and $B(y_i) \rightarrow E(y_i)$. 

We have studied the rate asymmetries over
much of the parameter space in this model and have
found the PRA for $b \to s \gamma$ to be very small, consistent
with our general observations above.
For $b\to d \gamma$, the PRA shows a strong dependence on
the CKM phase (see Eq.~(\ref{eq:acpbdgappr}))
but is barely affected by the new phase $\delta$.
Interestingly, the CKM phase can take a wide range of values in this model
because the observed CP violation in the neutral kaon system
may come partly or solely from the charged Higgs sector~\cite{ksw}.
As a result, the PRA for $b\to d \gamma$ can be very different from
the SM expectation, and can even be quite small.
These features are illustrated in 
Figs.~\ref{fig:2ha} and~\ref{fig:2hb}.
The shaded regions in these plots indicate the ``allowed'' regions,
while the white regions are ruled out by constraints coming
from $\epsilon_K$, $\Delta m_K$ and the branching ratio
for $b \to s \gamma$.  The constraints coming from $\epsilon_K$ 
and $\Delta m_K$ were discussed in
detail in Ref.~\cite{ksw} for several scenarios
and have been adopted here without change for similar scenarios.
Figure~\ref{fig:2ha} illustrates the case of a SM-like scenario:
the Wolfenstein parameters $\rho$ and $\eta$ take their SM best
fit values, and the CP asymmetry for $b\to d \gamma$ varies
in the neighborhood of its SM prediction.
Figure~\ref{fig:2hb} shows a real-CKM  scenario, where
$\eta=0$ and $\rho=0.42$.  In this case
the CP asymmetry for $b\to d \gamma$  is close to zero.
Note that the recent CDF measurement of the CP asymmetry in 
$B\to \psi K_S$~\cite{cdf} disfavours a negative value for
$\eta$, both within the SM and within the T2HDM~\cite{ksw}.
We thus estimate the maximum CP asymmetry for $b\to d\gamma$
within the T2HDM to be of order $+4\%$, as indicated in 
Fig.~\ref{fig:2hb}.

It is worthwhile to compare
the rate asymmetries in the 3HDM and
the T2HDM.  It is evident from Figs.~\ref{fig:3ha}-\ref{fig:2hb} that
the rate asymmetry for $b \to s \gamma$ is 
small in both models, as argued above on general grounds.
In contrast, the rate asymmetry for $b \to d \gamma$ in these models
can differ appreciably from its SM value, $-16\%$, taking on values between $\pm 20\%$
in the 3HDM and between approximately $-16 \%$ and $+4\%$ in the T2HDM.
An important difference between the two models is that $A_{CP}^{b\to d\gamma}$
in the T2HDM is almost completely determined by the CKM phase, while
the analogous asymmetry for the 3HDM exhibits a strong dependence on 
the charged Higgs sector CP-violating phase.

\section{Supersymmetry}
\label{sec:susy}

Supersymmetric predictions for the $b \to s \gamma$ rate asymmetry
are highly model-dependent
 \cite{kagneub2,kagneub,goto,aoki,chua,kim,baek,giusti,chun}. 
 Restrictive models, such as minimal supergravity,
yield asymmetries of order $2\%$ or less when the electron
and neutron electric dipole moment constraints are taken into 
account~\cite{goto}.  Relaxing some
of the assumptions of minimal supergravity leads to asymmetries
as large as $10\%$~\cite{aoki}, while allowing for non-CKM-like intergenerational
squark mixing can lead to asymmetries of order $15\%$~\cite{chua,kim} or larger 
if the gluino mass is significantly lighter than the squark masses \cite{kagneub}.
In this latter case, gluino-mediated diagrams become important in addition to
the usual chargino- and charged Higgs-mediated diagrams.  Allowing
for the violation of $R$-parity can lead to asymmetries of order $17\%$~\cite{chun}.
A recent calculation of the $b \to d \gamma$ rate asymmetry
in a fairly restrictive supersymmetric scenario yielded
values in the ranges $-(5-45)\%$ and $+(2-21)\%$~\cite{asatdgamma}.

The $b \to s \gamma$ rate asymmetry within supersymmetric models can
be appreciably larger than that in Higgs models or in the SM.  It
is instructive to see how supersymmetric models evade the accidental
cancellation encountered by Higgs models.
In order to understand this, let us
consider the chargino-squark loop contributions that must
typically be added to the
$W$ and charged Higgs contributions considered thus far. These diagrams
give rise to additional terms in the Wilson 
coefficients~\cite{aokinote},
which we denote by
\begin{eqnarray}
C_7^{\omega}(m_W)  & = & f_CK^7_1(y) + f_D K^7_2(y) \; ,\\
C_8^{\omega}(m_W)  & = & f_CK^8_1(y) + f_D K^8_2(y) \; .
\end{eqnarray}
Here $f_C$ and $f_D$ parameterize the couplings and $y$ is the square
of the ratio of the chargino and squark masses.
The chargino loop for $C_7$ involves two new  Inami-Lim functions,
$K^7_1 =6A(y)+5D(y)$ and $K^7_2=6B(y)+5E(y)$,
whereas for $C_8$ the two functions are given by
$K^8_1=9A(y)-6D(y)$ and $K^8_2=9B(y)-6E(y)$.
Since $D/A\sim 1$ and $E/B\sim 1$, we have $K^8_1/K^7_1 \sim 3/11$ and 
$K^8_2/K^7_2 \sim 3/11$.
As a result, $C_7^{\omega}(m_W)$ is generically a few times larger
than $C_8^{\omega}(m_W)$, and there is no longer a cancellation
between the first two terms of Eq.~(\ref{acpgen3}).
Thus, a large asymmetry in supersymmetric theories is possible.

\section{Comments on experimental techniques and sensitivities}
\label{sec:exptech}

The experimental detection of the inclusive process $B\to X_d \gamma$ (i.e.,
$b\to d\gamma$) is very challenging. The main difficulty is that the
smaller signal for $b\to d\gamma$ must be isolated from the much larger
``background'' of $b\to s \gamma$. This requires ensuring that the
signal events for $b\to d\gamma$ do not carry net strangeness. In the following
we discuss two ideas that may be used to determine
the branching ratio and partial rate asymmetry for $B\to X_d \gamma$.
We also compare the experimental sensitivity to the PRA's for 
$b \to s \gamma$ and $b\to d \gamma$.

\subsection{Multiplicity cut}

One strategy that may be helpful 
in determining ${\cal B}(B \to X_d\gamma)$ 
is to use a multiplicity cut so that one
deals with a semi-inclusive sample consisting of a maximum of $n$ (say 5)
mesons. Using such a measurement along with the corresponding
one for $B\to X_s\gamma$, one can deduce the {\em total} inclusive rate for
$B\to X_d\gamma$ to a very good approximation via:
\begin{eqnarray}
	\!\!\!\!\frac{{\cal B}(B\to X_d \gamma)}{{\cal B}(B\to X_s \gamma)} 
		& & \nonumber \\
		& \hspace{-1.3in}\approx & \hspace{-.5in}\frac{\Gamma(B\to
		\gamma+2\pi) +\Gamma(B\to\gamma+3\pi) +\cdots+\Gamma(B\to\gamma
		+n\pi)}{\Gamma(B\to \gamma+K+\pi) +\Gamma(B\to\gamma+K+2\pi)
		+\cdots+\Gamma( B\to\gamma+K + (n-1)\pi)} \label{bapproxgam}
\end{eqnarray}
Note that each value of $n$ for which a measurement is available leads
to such an equation.  Thus, in principle,
the measurements could be repeated for different values of $n$ to help
reduce systematic uncertainties~\cite{note:kstar}.
There are two key assumptions used in Eq.~(\ref{bapproxgam}). The first
is that $SU(3)$ holds; this assumption could introduce errors
at the $10\%$ level.
The second assumption is that 
non-spectator contributions may be ignored -- i.e., Eq.~(\ref{bapproxgam})
assumes that the spectator
approximation holds precisely.  This latter assumption should work very well 
for neutral $B$'s. For charged $B$'s, the annihilation graph can make a
contribution of order 20\%~\cite{abs}, so that Eq.~(\ref{bapproxgam})
will receive somewhat
larger corrections for $B^\pm$; nevertheless Eq.~(\ref{bapproxgam})
provides an excellent basis for the determination of an
approximate expression for ${\cal B}(B\to X_d \gamma)$.

\subsection{End point study via an energy cut}

Another possibility is to place a mass cut on the recoiling hadron so
that $m_X<(m_\pi+m_K)\sim 635$ MeV.  Correspondingly there is a minimum
energy cut on the photon, e.g., $E_\gamma \gsim 2.60$ GeV~\cite{note:endpt}.
 This would
ensure that the decay products result from $b\to d \gamma$ and not from
$b\to s\gamma$. PRA determination from this would be quite
straightforward:
\begin{eqnarray}
\left. A^{b\to d\gamma}_{CP}\right|_{E_\gamma>E^{\rm min}_\gamma} =
	\frac{{\cal B}^\prime(\overline{B}\to X_d\gamma)-
		{\cal B}^\prime(B\to X_{\overline{d}}\gamma)}
	{{\cal B}^\prime(\overline{B}\to X_{d}\gamma) + 
		{\cal B}^\prime(B\to X_{\overline{d}}\gamma)} \; ,
\end{eqnarray}
where ${\cal B}^\prime$ stands for the branching ratio with the energy cut on
the photon.

\subsection{PRA in $b\to d \gamma$}
\label{sec:prainbdgamma}

Since the number of $B$'s needed for searching for a given asymmetry $A_{CP}$ scales
approximately as $({\cal B} A_{CP}^2)^{-1}$, the PRA in $b\to d\gamma$ is
statistically much more accessible than that in $b\to s\gamma$. The
point is that, based on the SM, the PRA in $b\to d\gamma$ is expected
to be 20-40 times larger than that in $b\to s\gamma$. Not only does this
offset the difference in their branching ratios, it may even overcome the decrease in
detection efficiency for $b \to d\gamma$.

As an illustration, suppose ${\cal B}(B\to X_d\gamma)\sim {\cal B}(B\to X_s\gamma)
/20\sim 1.5\times10^{-5}$. Then, to establish a 16\% PRA to
$3 \sigma$ significance with a 20\% detection efficiency and a $25\%$
tagging efficiency, the needed number of
$B$-$\overline{B}$ pairs is:
\begin{eqnarray}
	N^{3\sigma}_B \sim \frac{9}{1.5\times10^{-5}\times0.2\times 0.25\times
	0.16\times0.16} = 4.7\times10^8
\end{eqnarray}
In contrast, let us take $A^{b\to s \gamma}_{CP}\simeq 0.6\%$,
a detection efficiency of 70\% and a 50\% tagging efficiency.
In this case the number of $B$-$\overline{B}$ 
pairs needed is:
\begin{eqnarray}
	N^{3\sigma}_B \sim
	\frac{9}{3.2\times10^{-4}\times0.7\times 0.50 \times0.006\times0.006} =
	2.2\times10^9
\end{eqnarray}
Therefore, establishing a PRA in $b \to d \gamma$ may be less difficult
than in $b \to s\gamma$.

\section{Concluding remarks}
\label{sec:concl}

    The study of CP violation in $B$ decays is one of the 
central themes at the various $B$-facilities.  In particular, 
the standard model CKM paradigm of CP violation will be tested
and new sources of CP violation will be searched for.
In this work, we have shown that the charged Higgs loop contribution 
to the rate asymmetry in $b \to s \gamma$ is small in all models
due to an accidental cancellation in the loop integrals. 
More specifically, in charged Higgs models, $C_8^{\rm new}\sim C_7^{\rm new}$.
As a result, the first two terms in Eq.~(\ref{acpgen3}) nearly cancel
and the third term is suppressed, leading to a small asymmetry.
Therefore, any experimental discovery of direct CP violation 
in $b \to s \gamma$ at the $10\%$ level will be due
to new sources of CP violation beyond charged Higgs, for example
supersymmetry.
We have also explained why chargino loops in supersymmetry 
do not undergo this cancellation.  This is because
$C_7^{\omega}(m_W)$ is generically a few times larger
than $C_8^{\omega}(m_W)$.  Consequently,
sizeable asymmetries are possible within supersymmetry.

The direct CP asymmetry in $b \to d \gamma$ behaves quite
differently from its counterpart in $b\to s\gamma$.
In Multi-Higgs doublet models,
the asymmetry can be large and of opposite sign relative to the SM.
In the T2HDM, in particular,
CP violation is mainly determined by the CKM phase, and
the asymmetry can show a strong deviation from the SM prediction
if the observed CP violation in the neutral kaon system 
has a sizable component from the charged Higgs sector. 
These features make the the experimental measurement of 
the rate asymmetry in $b\to d \gamma$ very worthwhile.
Our main results are summarized in Table~\ref{tableone} along with representative
results from various other models.

 We have also offered some suggestions that we hope would lead to 
the measurement of the inclusive decay ${\cal B} (B \to X_d \gamma)$, 
and eventually even the PRA in that mode.

\acknowledgments
This research was supported in part by the U.S. Department of Energy
contract numbers  DE-AC02-98CH10886 (BNL) and
DE-FG03-96ER40969 (Oregon).
K.K. is supported by an award from Research Corporation and by
the SRTP at Taylor University.
G.W. would like to thank the BNL theory group for its hospitality 
during the final stage of this work.

\newpage

\begin{table}
\begin{tabular}{|c||c|c|} 
	~~~~~~~~~~~~Model~~~~~~~~~~~~ & ~~~~~~~$A_{CP}^{b\to s \gamma}~(\%)$~~~~~~~ 
		& ~~~~~~~$A_{CP}^{b\to d \gamma}~(\%)$~~~~~~~ \\ 
		\hline\hline
	SM & 0.6 & $-16$ \\ \hline
	2HDM (Model II) & $\sim 0.6$ & $\sim -16$ \\ \hline
	3HDM & $-3$ to $+3$ & $-20$ to $+20$ \\ \hline
	T2HDM & $\sim 0$ to $+0.6$ & $\sim -16$ to $+4$ \\ \hline
	Supergravity~\cite{goto,aoki,asatdgamma} & $\sim -10$ to $+ 10$ & $-(5-45)$ and $+(2-21)$
			\\ \hline	
	~~~SUSY with squark mixing~\cite{chua,kim,kagneub}~~~ & $\sim -15$ 
 to $+15$ & \\ \hline
	SUSY with $R$-parity violation~\cite{chun} & $\sim -17$ to $+17$ & \\
\end{tabular}
\vspace{0.2in}
\caption{CP asymmetries in various models.  Quoted ranges are approximate; see the text for details.
The blank entries represent quantities that have not, to our knowledge, been considered in the literature.
The rate asymmetry for $b\to s \gamma$ from SUSY with squark mixing could be larger than the quoted
$\pm 15\%$ if the gluino mass is significantly lighter than the squark masses [9].
Note that the range quoted for the $R$-parity violating case assumes $m_{\nu_\tau} \sim 10$~keV.
The asymmetry is negligible if $m_{\nu_\tau}\ll 10$~keV, as is indicated by the SuperKamiokande 
atmospheric neutrino oscillation data [52].}
\label{tableone}
\end{table}

\begin{figure}[hbt]
\epsfxsize5in
\epsfbox{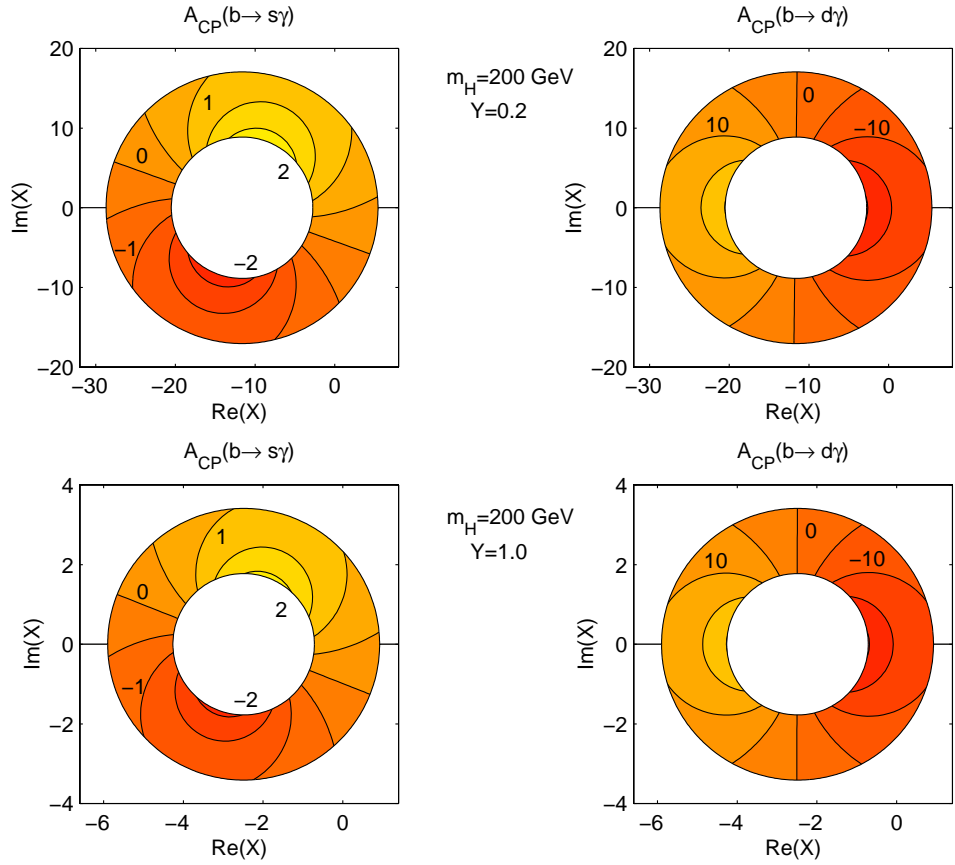}
\medskip
\medskip
\caption{Same as Fig.~\ref{fig:3ha}, but with different values
for $m_H$ and $Y$.  }
\label{fig:3hb}
\end{figure}

\begin{figure}[hbt]
\epsfxsize5in
\epsfbox{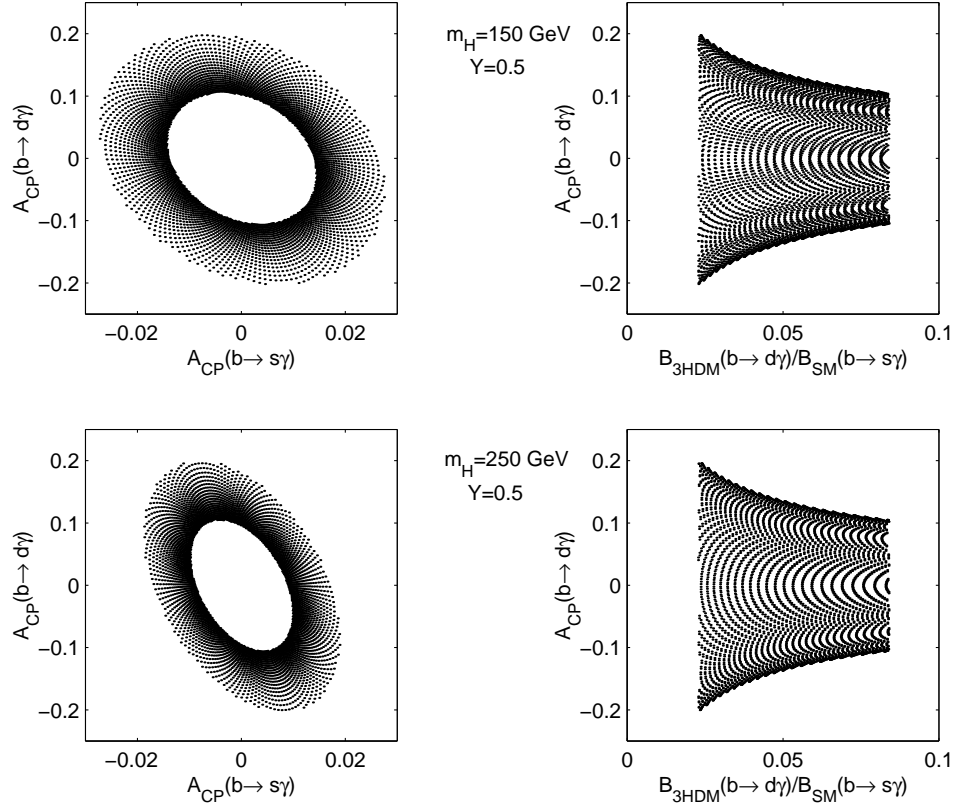}
\medskip
\medskip
\caption{Correlation between the rate asymmetries
of $b\to s \gamma$ and $b\to d \gamma$ in the 3HDM. 
Also shown are the ranges of the ratio of  branching ratios
for the radiative decays. }
\label{fig:3hc}
\end{figure}

\begin{figure}[hbt]
\epsfxsize5in
\epsfbox{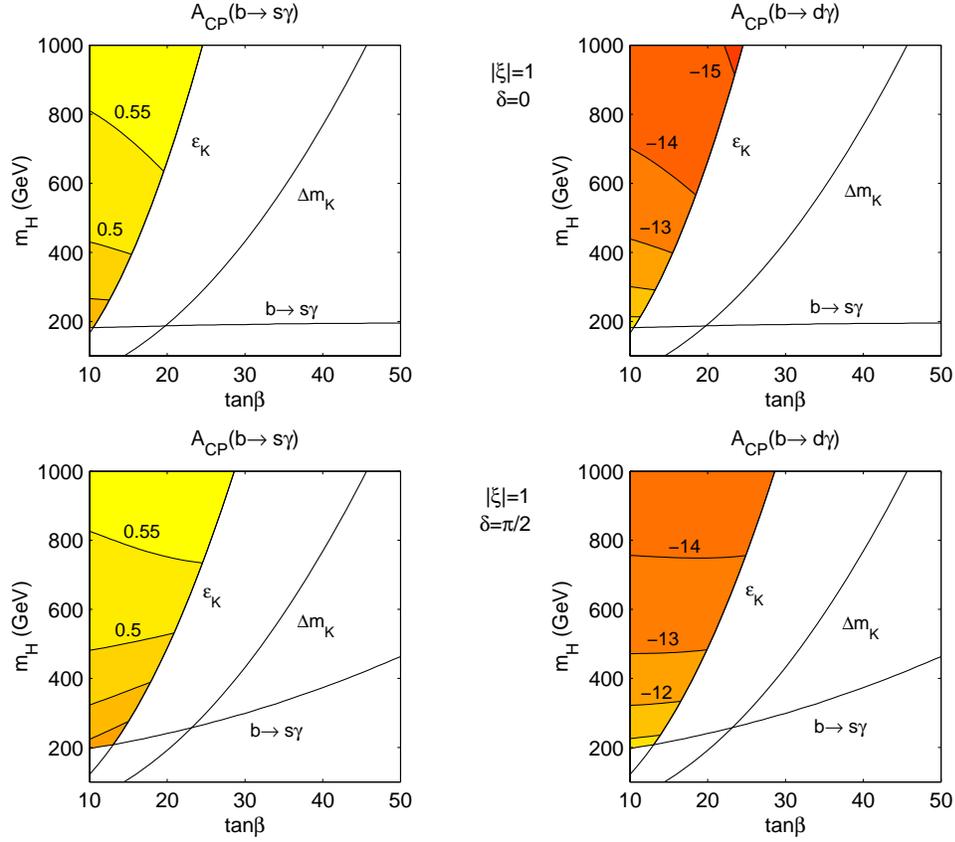}
\medskip
\medskip
\caption{Contour plots of PRA's in $b\to s \gamma$ and
$b \to d \gamma$  in the T2HDM -- SM-like scenario:
 $\rho$ and $\eta$ take their SM best fit values 
($\rho=0.20$ and $\eta=0.37$).
The experimentally allowed regions are shaded, and 
the contours show the asymmetries as percentages.}
\label{fig:2ha}
\end{figure}

\begin{figure}[hbt]
\epsfxsize5in
\epsfbox{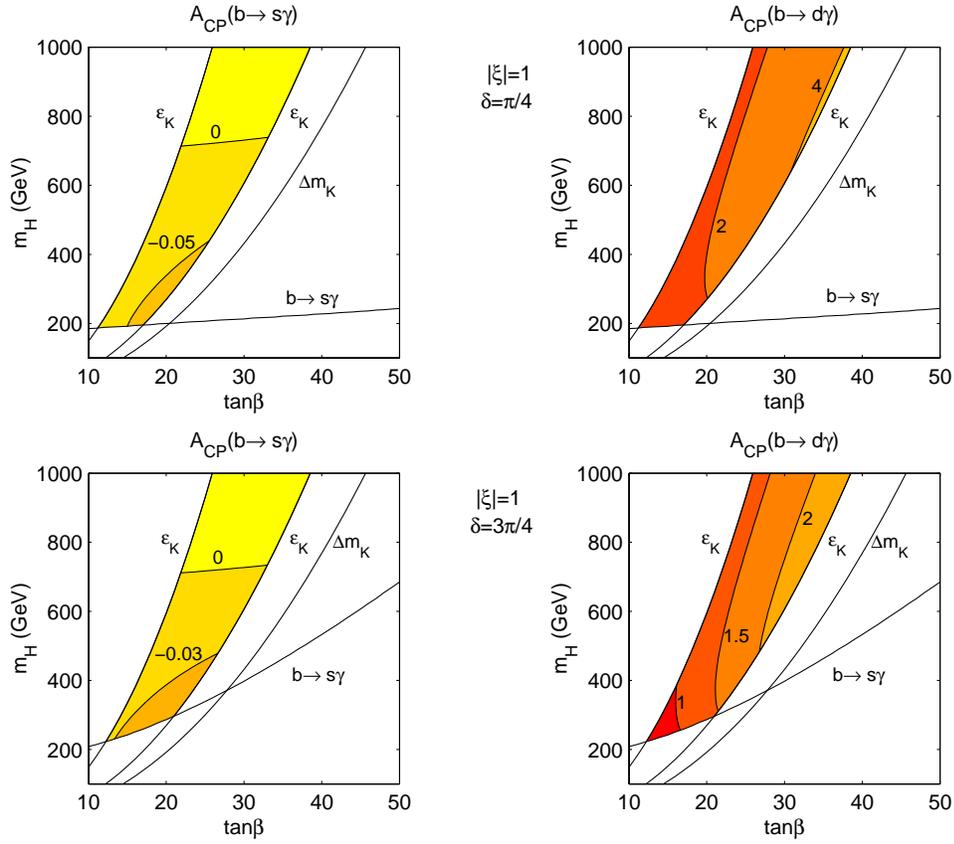}
\medskip
\medskip
\caption{Similar to Fig.~\ref{fig:2ha}, but with a real CKM matrix. }
\label{fig:2hb}
\end{figure}


\begin{thebibliography}{99}
\bibitem{c7note}
	Note that $C_{7,8}(m_b)$ are often denoted by $C_{7,8}^{\rm eff}(m_b)$
	in the literature.
\bibitem{note:beta_CKM}
	We add the subscript to $\beta_{CKM}$ in order to distinguish it from
	$\beta\equiv \tan^{-1}v_2/v_1$.
\bibitem{fcnhnote}
	In models that do not enforce NFC there can also be
	flavour-changing neutral Higgs interactions. 
	Also, in the T2HDM (to be discussed below),
	there are contributions from the charm quark that should be taken into
	account.
\bibitem{lightnote}
	We only consider effects due to the lighter of the two physical charged Higgs.
\bibitem{zetanote}
	If $X\approx 0$ the $|Y|^2$ terms in Eq.~(\ref{zeta3hdm})
	dominate and $\zeta\simeq D/A -1$, which can be
	relatively large for small Higgs masses (see Fig.~\ref{fig:inami}(b)). 
	For $m_H=150$~GeV and $X=0$, for example, $|\zeta|\simeq 0.42$.
	The asymmetry is still very small, however,
	due to the smallness of $Y$ and to the factor of ``$1/6$'' appearing
	in Eq.~(\ref{c7c83hdm}).  Moderate values of $\zeta$ can also occur if the
	$XY^*$ terms dominate and $m_H$ is very large; however, large Higgs masses 
	tend to suppress the nonstandard contributions to $A_{CP}^{b\to s \gamma}$.
\bibitem{ratebsg}
	See Refs.~\cite{bbl} and~\cite{czarmarc}.  We use $\alpha_{\rm em}$, as 
	suggested in Ref.~\cite{czarmarc}.
\bibitem{transnote}
	It is straightforward to show that the 
	transformation on $X_R$ and $X_I$ leads to a shift and rescaling that
	leaves the ``shapes'' of the contours and that of the circular allowed
	region invariant with respect to each other.
\bibitem{note:kstar}
	In using Eq.~(\ref{bapproxgam}), contributions from resonance channels
	(such as $B\to K^* \gamma$, $B\to \rho \gamma$,\ldots) can easily be
	excluded or included, by using the final states into which the resonances
	decay, as long as the numerator and denominator are treated in the
	same way.
\bibitem{note:endpt}
Once again from a measurement of the partial width in such a kinematic 
window, we can deduce the branching ratio by appealing to $SU(3)$:
$ {\cal B} ( B \to X_d \gamma)/ {\cal B} ( B \to X_s \gamma)
\simeq  {\cal B}^\prime ( B \to X_d \gamma)/ {\cal B}^\prime
 ( B \to X_s \gamma)$, where the photon energy cuts for the two decay
modes have the same fraction of their respective available energies.
\end{thebibliography}
\end{document}